\documentclass[aps,prd,preprintnumbers,nofootinbib,floatfix,floats,superscriptaddress]{revtex4}
\usepackage{bm}
\usepackage[utf8]{inputenc} 
\usepackage{latexsym}
\usepackage{dcolumn}
\usepackage{amsmath,amsfonts,amssymb,mathtools}
\usepackage{graphicx,epsfig}
\usepackage{float}
\usepackage{color}
\usepackage[active]{srcltx}%DO NOT DELETE /
\usepackage[caption=false]{subfig}
\usepackage{slashed}

\usepackage{comment}
\usepackage{tikz}
\usepackage{multirow}
\usepackage[normalem]{ulem}%for strikethrough
\usepackage[mathscr]{euscript}
\usepackage{mathrsfs}
\usepackage{enumitem}
\usepackage{booktabs}
\usepackage{array}
\usepackage{multirow}
\usepackage{xcolor}
\usepackage{hyperref}

\newcommand{\gae}{\lower 3pt \hbox{$\,\, \buildrel {\scriptstyle >}\over {\scriptstyle
\sim}\,\,$}}
\newcommand{\lae}{\lower 2pt \hbox{$\, \buildrel {\scriptstyle <}\over {\scriptstyle
\sim}\,$}}

\DeclarePairedDelimiterX\braket[2]{\langle}{\rangle}{#1\,\delimsize\vert\,\mathopen{}#2}
\newcommand{\nobarfrac}{\genfrac{}{}{0pt}{}}

\begin{document}

%\begin{comment}

\title{Geodesic structure of spacetime near singularities}

\author{Mayank}
\email{mayank@physics.iitm.ac.in}
\affiliation{Centre for Strings, Gravitation and Cosmology, Department of Physics, Indian Institute of Technology Madras, Chennai 600 036, India}

\author{Dawood Kothawala}
\email{dawood@iitm.ac.in}
\affiliation{Centre for Strings, Gravitation and Cosmology, Department of Physics, Indian Institute of Technology Madras, Chennai 600 036, India}

\date{\today}
\begin{abstract}
\noindent Geodesic flows emanating from an arbitrary point $\mathscr{P}$ in a manifold $\mathscr{M}$ carry important information about the geometric properties of $\mathscr{M}$. These flows are characterized by Synge's world function and van Vleck determinant - important bi-scalars that also characterize quantum description of physical systems in $\mathscr{M}$. If $\mathscr{P}$ is a regular point, these bi-scalars have well known expansions around their flat space expressions, quantifying \textit{local flatness} and the equivalence principle. We show that, if 
$\mathscr{P}$ is a singular point, the scaling behavior of these bi-scalars changes drastically, capturing the non-trivial structure of geodesic flows near singularities. This yields remarkable insights into classical structure of spacetime singularities and provides useful tool to study their quantum structure.
\end{abstract}

\maketitle

\tableofcontents

%%%%%%%%%%%%%%%%%%%%%%%%%%%%%%%%%%%%%%%%%%%%%%%%%%%%%%%%%%%%%%%%%%
\section{Introduction} \label{sec:intro}

An important first step in understanding the role of quantum effects in regions of strong gravitational fields is to understand the behavior of geodesic flows - the behavior of geodesics emanating from a single point - in vicinity of the singularity. This is so because many 
important features of classical and quantum dynamics depend on this behavior; see, for instance, \cite{Gutzwiller1990} for an extensive textbook discussion on this 
point. While the structure of these flows is well understood in geodesically convex neighborhoods of regular spacetime points, not much is known about their 
behavior in regions containing curvature singularities. This is to be expected since covariant Taylor expansions that work well in the former case breaks down in 
the latter, and hence the analysis of geodesic flows directly in terms of curvature-related quantities can not be done for generic singularities. We derive here, 
for the first time, explicit behavior of Synge's world function $\Omega(x, y)$ and van Vleck determinant $\Delta(x, y)$ - two important bi-scalars characterizing geodesic flows - in the vicinity 
of FLRW and Kasner-like singularities. In the process, we improve upon an old result by Buchdahl \cite{Buchdahl:1972rw}\cite{Buchdahl:1980iml} who calculated the series expansion of ``$\sqrt{|\Omega|}$" in FLRW and obtained a series whose terms can be shown to diverge as one approaches the singularity at $t \to 0$. This obviously un-physical result, as we shall see, is an artifact of expanding a wrong function (and manifests, in fact, even in the Minkowski limit!), and was presumably not noticed by Buchdahl. Here, we focus instead on $\Omega(x, y)$ itself, which we take to be the only meaningful quantity.  
\\ \\
\noindent The results exhibit a remarkable and non-trivial modification in the behavior of these bi-scalars near the singularities. We elaborate on these results and put them in the broader context of quantum gravity. Over the past decade, several research works on quantum spacetime have recognized the relevance of Synge's world function and van Vleck determinant in reconstructing an ``effective" spacetime metric $q_{ab}(x; x_0)$ \cite{Kothawala:2013maa},\cite{Kothawala:2023tuh},\cite{qmetric-rev1}, \cite{Padmanabhan:1996ap}, \cite{Pesci:2025ttg}. The effective metric description already yields some very interesting insights into local properties of spacetime in a geodesically convex neighborhood of a regular event $\mathscr{P}_0$. Our results here enable extending these insights when $\mathscr{P}_0$ represents a curvature singularity. These results will be presented in a separate work \cite{DK2026} which, in fact, provided an important motivation for the current work.
\\ \\
\noindent To keep our discussion streamlined and focus clear, we give definitions of the key quantities and summarize our key results at the end of this section (see below). Rest of the paper will provide a detailed exposition on various aspects of the results, and is organized as follows: Section II gives the key derivation leading to systematic representations of $\Omega(x,y)$ and $\Delta(x,y)$ in FLRW and Bianchi spacetimes. In Section III, we use these results to discuss the geodesic and causal structure of these spacetimes, focussing on the limit where the base point is taken as the singularity as $t \to 0$. Finally, we conclude with a brief discussion in Section IV.

\subsection{Synge's world function and van Vleck determinant} 

Synge's world function and van Vleck determinant are two important bi-scalars that characterize measurements in curved spacetimes, and also appear ubiquitously in quantum field theory. Two-point (and higher) correlators of quantum fields in curved spacetime have a universal expansion in terms of the world function near the coincidence limit. These expansions form the backbone of the point-splitting program for regularization in curved spacetime, and have been developed and employed extensively by de Witt \cite{DeWitt:1975}, Cristensen \cite{Christensen:1976},\cite{Christensen:1978}, Parker\cite{ParkerFulling:1974} and others in deriving covariant expansions of effective lagrangians in curved spacetimes, covariant formulation of path integral etc. The use of the world function as a fundamental object to study spacetime dates back to Synge \cite{Synge:1960}, where it is defined operationally as half the square of the geodesic distance between the two spacetime points (denoted by $\Omega(x,x')$). The spacetime metric $g_{ab}(x)$ can be constructed as the coincidence limit of $\nabla_{a}\nabla_{b}\Omega(x,y)$, i.e.

\begin{eqnarray}
    g_{ab} = \lim_{x'\longrightarrow x}\nabla_{a}\nabla_{b}\Omega(x,x') = [\nabla_{a}\nabla_{b}\Omega(x,x')]
\end{eqnarray}

Similarly, higher order coincidences gives us information about the spacetime curvature. Thus world function can be used to study tensor expansions on a given spacetime manifold in a covariant manner. The only hindrance to its use (apart from the possible existence of multiple geodesics connecting two points) is the non-existence of any closed-form expression for the geodesic interval as a function of spacetime coordinates, even in highly symmetric spacetimes. In fact, to the best of our knowledge, closed-form expressions are known only in maximally symmetric and plane-wave spacetimes \cite{hari:2021gns},\cite{Harte:2012uw}. It would surprise the in-cognoscenti that the simplest and most widely encountered spacetimes, such as Schwarzschild and FLRW, have no closed-form expression for 
$\Omega$. In this work, we derive a systematic representation for the world function in FLRW and Bianchi spacetimes, in a form that is applicable near singularities in these spacetimes.   

 \begin{figure*}[htbp]%
                \centering
                {{\includegraphics[width=0.8\textwidth]{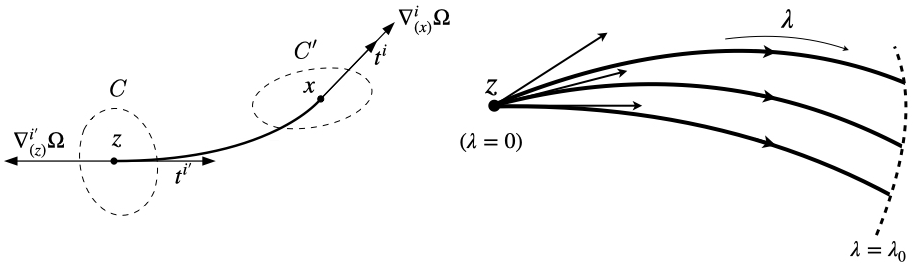} }}
                \caption{The most basic non-local observables in an arbitrary curved manifold are Synge's world function and the van Vleck determinant. The former characterizes single geodesics (\textbf{left}), while the latter measures the \textit{density of geodesics} emanating from a given point (\textbf{right}). $\lambda$ is the affine parameter along the geodesic. See text for more details.}
            \label{fig:geodesic-spray}%
 \end{figure*}

Another important bi-tensor of central importance that can be computed from the world function directly is the van Vleck determinant. It appears at various places, like fluctuation factor in semi-classical approximation to path integrals in curved spacetime \cite{BekensteinParker:1981}, heat kernel expansions in curved spacetime \cite{Avramidi:2015}, reconstruction of spacetime from two-point correlators \cite{Sorkin:2017},\cite{Padmanabhan:2020},\cite{KothawalaPadmanabhan:2014}, etc. Given a geodesic with endpoints $x$ and $z$, the van Vleck determinant is defined as,

\begin{eqnarray}\label{eq:VVD_def}
  \Delta(x,z) := -\frac{1}{\sqrt{-g(x)}}\det\left( \frac{\partial^{2}\Omega(x,z)}{\partial x^{a}\partial z^{b}} \right)\frac{1}{\sqrt{-g(z)}}    
\end{eqnarray}
Intuitively, $\Delta(x,z)$ gives a quantitative measure of \emph{density of geodesics} emanating from a given point $z$ and measured at points $x$ a fixed geodesic distance from $z$; this is depicted in Fig. \ref{fig:geodesic-spray}.

\begin{comment}
    \begin{figure}
    \centering
    \includegraphics[width=0.5\linewidth]{fig_1.pdf}
    \caption{The anchor point $\mathscr{P}$ is a cosmological singularity with $x^{a}$ and $X^{a}$ to be any two generic points on a time-like geodesic.}
    \label{fig:generic_setup}
\end{figure}
\end{comment}

\subsection{Summary of results} 

Since our analysis will be somewhat involved, we summarize our main results in Table \ref{eq:summary_table}, which gives the coincidence and singularity limits for the various biscalars in FLRW and Bianchi Type I spacetime. The reader is requested to refer to the relevant section(s) for definitions of various quantities in the table, which we have excluded here to avoid clutter.
We will also present an infinite series representation for Synge's world function for FLRW and Kasner (Schwarzschild) spacetime. The limits of $\Delta$ and $\Box \Delta^{1/2}$ and $K=\textrm{Tr}(K_{ab})$ are used to study the geodesic and causal structure near the curvature singularity. The structure of light cones is illustrated in Fig. \ref{fig:equi-geodesic_surface} and Fig. \ref{fig:3Dkasner} by plotting null geodesics along with the $\Omega=$ constant surfaces for different (fixed) points in the neighborhood of FLRW and Kasner (Schwarzschild) singularity.

\begin{table}[htbp]\label{eq:summary_table}
\centering
\begin{tabular}{|c|c|c|c|c|}
\toprule
\hline
\textbf{Bitensor} & \multicolumn{2}{c|}{\textbf{FLRW}} & \multicolumn{2}{c|}{\textbf{Bianchi Type I (Schwarzschild)}} \\
\midrule
\hline
& \textbf{Coincidence} & \textbf{Singularity} & \textbf{Coincidence} & \textbf{Singularity} \\
\midrule
\parbox{2.1cm}{ 
\begin{align*}
&\\[1em]
&\Delta(t,T)
\end{align*}
} 
& 
\parbox{3.2cm}{
\begin{align*}
&\textrm{matter} :1 + \frac{\varepsilon^2}{9T^2} + \cdots \\[0.3em]
&\textrm{radiation}:1 + \frac{\varepsilon^2}{8T^2} + \cdots
\end{align*}
}
&
\parbox{4.2cm}{
\begin{align*}
& \textrm{matter}:-\frac{\ell^{2}t}{972\ T^{5/3}t_{0}^{4/3}} -\frac{\ell^{2}t}{243\ T^{4/3}t_{0}^{4/3}} \\& \ \ \ \ \ \  \ \ \  \ \ + \frac{t}{27T}\left( 1 - \frac{\ell^{2}(t/t_{0}^{4})^{1/3}}{4t} \right)\\[0.3em]
&\textrm{radiation}:\frac{\ell^{2}t^{3/2}(5-2\log(t/T))}{4\ T^{5/2} t_{0}(\log(t/T))^{6}} \\& \ \ \ \ \ \ \ \ \ + \left(\frac{t}{T}\right)^{3/2} \frac{1}{(\log(t/T))^3}
\end{align*}
}
&
\parbox{2.5cm}{
\begin{align*}
&1 + O(\varepsilon^4)
\end{align*}
}
&
\parbox{2.5cm}{
\begin{align*}
&-\frac{5\ell_{\perp}^{2}(t/t_{0}^{4})^{1/3}}{486\ T} \\&  -\frac{10\ell_{\perp}^{2}}{243\ T^{2/3} \ t_{0}^{4/3}} \\& + \frac{25t_{0}^{2/3}(z-Z)^{2}}{286\ t^{7/3} T^{1/3} \ t_{0}^{4/3}} \\& 
+ \frac{5t^{1/3}}{27 T^{1/3}}  + \frac{10}{27}\\& - \frac{25\ell_{\perp}^2}{243\ (t \ t_{0}^4)^{1/3}T^{1/3}}
\end{align*}
} \\
\cline{1-5}
\parbox{2.1cm}{
\begin{align*}
&\Box \Delta^{1/2}(t,T)\\[2.5em]
&K(t,T)
\end{align*}
}
& 
\parbox{3.2cm}{
\begin{align*}
& \textrm{matter}:\frac{2}{9T^{2}} \\[0.4em]
& \textrm{radiation}:\ \ 0\\[0.5em]
& \frac{3}{\varepsilon} + \frac{\ddot{a}}{a}\varepsilon
\end{align*}
}
&
\parbox{4.2cm}{
\begin{align*}
&\textrm{matter}:\frac{2267}{2187\sqrt{21}t^{2}}\\[0.3em]
&\textrm{radiation}: \ \ \  0\\[0.4em]
&\frac{3q}{t}
\end{align*}
}
&
\parbox{2.1cm}{
\begin{align*}
&\\[1em]
& \frac{3}{\varepsilon} - \frac{8\varepsilon^{3}}{1215T^{4}}
\end{align*}
}
&
\parbox{2.5cm}{
\begin{align*}
& \\[1em]
& \frac{8}{3t}
\end{align*}
} \\
\midrule
\hline
\parbox{2.1cm}{
\begin{align*}
& \Omega(t,T)
\end{align*}
}
&
\multicolumn{2}{c|}{
\parbox{7.4cm}{
\begin{align*}
&-\frac{ (t-T)^2}{2} + \sum_{j=1}^{\infty} \frac{\ell^{2j}}{(2j)!} 
\left[\frac{d^{2j-1}}{d\alpha^{2j-1}} \Bigl\{\frac{d\Omega(\alpha)}{d\alpha}(G)^{-2j}\Bigr\}\right]_{\alpha=0}
\end{align*}
}}
&
\multicolumn{2}{c|}{
\parbox{5cm}{
\begin{align*}
& \sum_{i,j,k=0}^{\infty}\mathcal{C}_{ijk}(x-X)^{i}(y-Y)^{j}(z-Z)^{k}
\end{align*}
}} \\
\hline
\bottomrule
\end{tabular}
\caption{Summary of important limits in FLRW and Bianchi Type I spacetimes}
\end{table}

The details of the results presented in Table \ref{eq:summary_table} will be discussed in the rest of the article. However, from the results summarized in the table above, we can immediately compare our series representation for $\Omega(t,T)$ for FLRW spacetime, with that given by Buchdahl for $\sqrt{|\Omega|}$, in the coincidence limit of small $\varepsilon$ and $\ell$, where $\varepsilon=t-T$ and $\ell=|\vec{x}-\vec{X}|$,

\begin{eqnarray}\label{eq:Omega_expansion}
    \Omega(t,T) \simeq -\frac{\varepsilon^2}{2} + \frac{\ell^{2}}{2} \Biggl(a(T)^{2}-a(T)\dot{a}(T)\ \varepsilon + \frac{1}{3}a(T)\dot{a}(T)\ \varepsilon^2 \Biggl) + \frac{\ell^{4}}{24} \Biggl( a(T)^2\dot{a}(T)^2  - a(T)\dot{a}(T)(\dot{a}(T)^2+ a(T)\ddot{a}(T))\varepsilon \Biggl)
\end{eqnarray}
which reduces to flat Minkowski result for $a(T)=\textrm{constant}$. On the other hand, the so-called Hamilton/characteristic function ``$\sqrt{|\Omega|}$" defined by Buchdahl in \cite{Buchdahl:1972rw}, gives divergences in coincidence limit at $\mathcal{O}(\ell^{4})$:
\begin{eqnarray}
    \sqrt{|\Omega|} \simeq - \frac{a(T)^4}{\varepsilon^3} + \textrm{less singular terms}
\end{eqnarray}

Our series representation \eqref{eq:Omega_expansion} is non-divergent both in the singularity limit $a(T) \to 0$ as well as in the coincidence limit $\varepsilon \to 0$, Thus $\Omega(t, T)$ serves as a more natural biscalar to study geometric properties of spacetime.
%%%%%%%%%%%%%%%%%%%%%%%%%%%%%%%%%%%%%%%%%%%%%%%%%%%%%%%%%%%%%%%%%%%
\section{Representations of 
\texorpdfstring{$\Omega$ and $\Delta$}{Omega and Delta} 
in FLRW and Bianchi spacetimes}

While the definition of the world function $\Omega(x,y)$ is straightforward, and it is numerically equal to one half of the squared geodesic interval between $x$ and $y$, it is extremely difficult to obtain a closed-form expression for it even in the simplest of spacetimes. The difficulty lies in determining the functional dependence of the $\Omega(x,y)$ on the end-point coordinates $x$ and $y$. It is due to this reason that one often has to resort to coincidence limit expansions of the world function in specific coordinates. However, such an expansion might be rendered useless simply due to the choice of bad coordinates, or near spacetime singularities, which is the case of interest for us in this work. We show that the world function in the vicinity of a class of singularities (FLRW or Kasner type) can be represented in an expansion that is mathematically well behaved.

\subsection{FLRW spacetime}

 In this section we consider the case of FLRW spacetime and compute the world function for a timelike geodesic. The FLRW metric with an arbitrary scale factor $a(t)$ is given by,
\begin{eqnarray}\label{eq:metric_FLRW}
    ds^{2} = -dt^{2} + a^{2}(t)(dx^{2} + dy^{2} + dz^{2})
\end{eqnarray}

 As mentioned in the Introduction, an attempt to calculate the series expansion of $\sqrt{\Omega}$ by solving the Hamilton-Jacobi equation, first made by Buchdahl \cite{Buchdahl:1980iml}, leads to spurious and unphysical divergences in the limit $t \to 0$. The only meaningful quantity to calculate is the Synge world function $\Omega$ itself, which, for a timelike geodesic, is given by, 
$\Omega=-\tau^{2}/2$, where $\tau$ is the proper time along the geodesic. A general approach to compute proper time as a function of end points amounts to solving the geodesic equation ($\nabla_{\bm u}\bm u=0$) for a timelike geodesic, which is in general a difficult task. We further restrict our construction to a normal convex neighborhood $\mathcal{N}$, 
within which any two points are connected by a unique geodesic lying entirely in $\mathcal{N}$. 
For a timelike congruence with initial expansion $\Theta_0 < 0$, the Raychaudhuri equation gives an upper bound on focusing time: $3/|\Theta_0|$ in four spacetime dimensions. Hence, provided the coordinate-time separation $\varepsilon = t - T$ between $(t,\vec{x})$ and $(T,\vec{X})$ satisfies $\varepsilon < 3/|\Theta_0|$, the connecting timelike geodesic segment remains free of caustics \cite{hetLam:2017wrp}. We will assume this to be the case. Moreover, owing to the presence of spacelike Killing vector fields in the FLRW spacetime, the computation of proper time can be reduced to a quadrature, yielding the following expressions \cite{Roberts:1993umg},

\begin{eqnarray}\label{eq:qudrature_eqns}
    \tau(t,T;\alpha) &=& \int_{T}^{t} \frac{a(t')}{\sqrt{\alpha^{2} + a^{2}(t')}}dt'\\
    \ell(t,T;\alpha) &=& {\alpha} \int_{T}^{t} \frac{dt'}{a(t')\sqrt{\alpha^{2} + a^{2}(t')}}= \alpha G(t,T;\alpha)
\end{eqnarray}

where $ \tau(t,T;\alpha)$ is the proper time as a function of coordinate time of end points and integration constant $\alpha$, $\ell(t,T;\alpha)$ is the geodesic distance on the three-space of constant curvature. Now $\Omega$ can be rewritten using ($d\tau^{2}/dt' = 2\tau (d\tau/dt')$) and we eliminate the integration constant $\alpha$ for $\ell$ using \textit{Lagrange-Good} formula \cite{Abdesselam:2002mj} in single variable to get $\tau(t,T;\ell)$ and arrive at the following expansion for $\Omega(t,T;\ell)$ (see Appendix \ref{FLRW_omega_derivation} for details),

\begin{eqnarray}
    \Omega(t,T;\ell) 
    %&=& \Omega(0)  + \sum_{j=1}^{\infty}\frac{\ell^{2j}}{(2j)!}\left[\frac{d^{2j-1}}{d\alpha^{2j-1}} (\alpha F)(G)^{-2j} \right]_{\alpha = 0}\\
    &:=& \Omega(0) + \sum_{j=1}^{\infty}\frac{\ell^{2j}}{(2j)!}\left[\frac{d^{2j-1}}{d\alpha^{2j-1}} \left[\frac{d\Omega}{d\alpha}(G)^{-2j} \right] \right]_{\alpha = 0}
\end{eqnarray}

where $\Omega(0)=-(t-T)^{2}/2$, This expansion in $\ell$ can be written in a closed form by identifying the following expansions, 

\begin{eqnarray}\label{eq:cosh_sinh}
        \cosh(\ell D_{\alpha} [(G)^{-1}\Omega(\alpha)])-1 &:=& \sum_{j=1}^{\infty}\frac{(\ell D_{\alpha})^{2j}[(G)^{-2j}\Omega(\alpha)]_{\alpha=0}}{(2j)!}\\
        \sinh(\ell D_{\alpha}[(G)^{-1}(G^{-1})'\Omega(\alpha)]) &:=& \sum_{j=1}^{\infty}\frac{(\ell D_{\alpha})^{2j-1}[(G)^{-(2j-1)}(G^{-1})'\Omega(\alpha)]_{\alpha=0}}{(2j-1)!}
\end{eqnarray}

where $D_{\alpha}=d/d\alpha$, so that the final expression for the world function can be written concisely as follows.

\begin{eqnarray}\label{eq:omega_flrw_concise}
     \Omega(t,T;\ell) = \Omega(0) + \cosh(\ell D_{\alpha}[G^{-1}\Omega(\alpha)])  + \ell \sinh(\ell D_{\alpha}[G^{-1}(G^{-1})'\Omega(\alpha)])
\end{eqnarray}

The Eq.\eqref{eq:omega_flrw_concise} is an analytic closed-form expression in terms of the world function $\ell$ on the three-space of constant curvature. The complication of further simplification arises because of the time-dependent scale factor $a(t)$, which is kept open for generality. The appearance of hyperbolic functions in the closed-form expression for $\Omega(t, T; \ell)$ provides a compact resummation of the underlying \emph{Lagrange-Good} series. This representation is analytic in $\ell, t, T$ provided the mapping $\ell = \alpha G(t, T; \alpha)$ remains locally invertible, a condition dictated by the non-vanishing of the Jacobian $d\ell/d\alpha$. Since $G(t,T;\alpha)$ possesses branch-point singularities in the complex plane at $\alpha = \pm i a(t')$ (see Eq.\ref{eq:qudrature_eqns}), the domain of validity for the world function is bounded by a maximum spatial distance $L_{\text{max}} = |\ell(i a(T))|$, where $a(T)<a(t)$ (expanding universe). The hyperbolic functions on the other hand elegantly package the higher-order derivatives of the series while ensuring that $\Omega(t, T; \ell)$ and the van-Vleck determinant remain analytic for all separations $\ell < L_{\text{max}}$, provided $a(t') \neq 0$ for $t' \in (t, T)$. The maximum spatial bound for FLRW spacetime comes out to be the following,

\begin{eqnarray}
    L_{\text{max}}(t, T) = 
\begin{cases} 
2\sqrt{t_0T} \ln\left( \frac{\sqrt{t} + \sqrt{t-T}}{\sqrt{T}} \right) & \text{Radiation} \\
3 t_0^{2/3}T^{1/3} \left( \frac{\sqrt{\pi} \Gamma(5/4)}{\Gamma(3/4)} - \left(\frac{T}{t}\right)^{1/3} {}_2F_1\left[\frac{1}{4}, \frac{1}{2}, \frac{5}{4}, \left(\frac{T}{t}\right)^{4/3}\right] \right) & \text{Matter}
\end{cases}
\end{eqnarray}

Even though the series representation now has a domain of validity $\mathscr{D}=\{\varepsilon<3/|\Theta_0|\ , \ \ell<L_{\textrm{max}}\}$, the hyperbolic functions can be used for analytical continuation of the series representation beyond branch points. As another self-consistency check, it can be verified that Eq.\eqref{eq:omega_flrw_concise} satisfies the HJ equation up to order $\mathcal{O}(\ell^{4})$ with;

\begin{eqnarray}\label{eq:sigma_order4}
    \Omega(t,T;\ell) = - \frac{(t-T)^{2}}{2} + \frac{(t-T)}{2I_{2}}\ell^{2} + \left[\frac{(t-T)I_{4} - \left( I_{2}\right)^{2}}{8\left( I_{2}\right)^{4}}\right]\ell^{4} + \mathcal{O}(\ell^{6})
\end{eqnarray}

which reduces to Minkowski limit for $a(t)=\textrm{constant}$,

\begin{eqnarray}\label{eq:sigma_flat_limit}
    \Omega_{\textrm{Flat}}(t,T;\ell)=\frac{1}{2}\left( -(t-T)^{2}+\ell^{2} \right)
\end{eqnarray}

where we have rescaled $\ell \to \sqrt{a}\ell$ for Minkowski limit. The Eq.\ref{eq:sigma_order4} is also the series expansion of $\Omega(t,T;\ell)$ connecting spacetime points $(t,\vec{x})$ to $(T,\vec{X})$ with expansion parameter $|\vec{x}-\vec{X}|=\ell$. The series converges $\forall \ \ (t,\vec{x}),(T,\vec{X}) \in \mathcal{N}$. Whereas $T=0$ is a branch point in $G(t,T;\alpha)$ corresponding to a curvature singularity, where we do an asymptotic expansion (\emph{singularity limit}) with $T\to 0$ and fixed $t$ to infer the behavior of various biscalars as we shall discuss later in this section.\\

It is worth emphasizing the generality of this result, since $\ell$ appearing in Eq.\eqref{eq:omega_flrw_concise} is the geodesic distance on the three-space of constant curvature. Thus, the expression for $\Omega(t,T;\ell)$ holds for flat, open and closed slicing of FLRW spacetime.\\ 

An immediate biscalar that we can construct from the $\Omega(t,T;\ell)$ is the van Vleck determinant $\Delta(t,T)$ which characterizes the geodesic spread about a given point, If we consider the expression \eqref{eq:sigma_order4} derived earlier, then the van Vleck determinant for FLRW spacetime up to $\mathcal{O}(\ell^{2})$ is given by, 

\begin{eqnarray}\label{eq:VVD_FLRW}
    \Delta (t,T) = \frac{1}{a(t)^{3}a(T)^{3}F(t,T)^{3}} + \frac{3(40F(t,T)^{3}Q(t,T) - F^{(1,1)}(t,T))}{2a(t)^{3}a(T)^{3}F(t,T)^{5}}\ell^{2} + \mathcal{O}(\ell^{4})
\end{eqnarray}

with 

\begin{eqnarray}\label{eq:F_and_Q}
    F(t,T) &=&  \frac{1}{t-T}I_{2}(a)\\\\
    Q(t,T) &=& \left[\frac{(t-T)I_{4}(a) -  I_{2}(a)^{2}}{8I_{2}(a)^{4}}\right]
\end{eqnarray}

where $I_{n}(a_{i})$ is defined by Eq.\eqref{eq:I_def}. As a consistency check, we explicitly verify in Appendix \ref{app:vvd_desitter} that this general expansion precisely reproduces the known exact analytical result for de Sitter spacetime perturbatively upto $\mathcal{O}(\ell^2)$. We now consider the limiting behaviour of $\Delta(t,T)$ along a timelike geodesic in coincidence and singularity limit in matter and radiation dominated FLRW spacetime. 
The expansion in $\epsilon=t-T<<T$ (\emph{coincidence limit}) and (\emph{Singularity limit}) $T<<t,T\to 0$ are obtained as follows,

\begin{enumerate}\label{eq:Delta_matter}
\item Matter-dominated FLRW spacetime 
   \begin{eqnarray}\label{eq:Delta_FLRW}
 \textrm{Coincidence}: \ \ \ \     \Delta(t,T) &=& 1 + \frac{\epsilon^{2}}{9T^{2}} + \mathcal{O}(\epsilon^{3},\ell^2)  \\
   \textrm{Singularity}: \ \ \ \ \Delta(t,T) &=& -\frac{\ell^{2}t}{972\ T^{5/3}t_{0}^{4/3}} -\frac{\ell^{2}t}{243\ T^{4/3}t_{0}^{4/3}} + \frac{t}{27T}\left( 1 - \frac{\ell^{2}(t/t_{0}^{4})^{1/3}}{4t} \right) +. . .  
\end{eqnarray}   
\item Radiation-dominated FLRW spacetime
\begin{eqnarray}\label{eq:Delta_radiation}
\textrm{Coincidence}: \ \ \ \  \Delta(t,T) &=& 1  + \frac{\epsilon^{2}}{8T^{2}} +\mathcal{O}(\epsilon^{3},\ell^2) \\    
 \textrm{Singularity}: \ \ \ \   \Delta(t,T) &=& \frac{\ell^{2}t^{3/2}(5-2\log(t/T))}{4\ T^{5/2} t_{0}(\log(t/T))^{6}} + \left(\frac{t}{T}\right)^{3/2} \frac{1}{(\log(t/T))^3}+...  
\end{eqnarray}
\end{enumerate}

The different divergent terms appearing in singularity limits are akin to what happens near caustics \cite{Harte:2012uw}. Note that $T=0$ defines a spacelike singularity  with $\ell<L_{\textrm{max}}(t,T\to 0)$ in the singularity limit expansion. This gives $\ell^2 \lesssim T$ for radiation and $\ell^2 \lesssim T^{2/3}$ for matter dominated spacetime. This indicates that the $\ell$-independent terms provide the dominant contribution in these singularity expansions. It can be independently verified that coincidence limit expansion for matter and radiation dominated FLRW spacetime agrees with, $\Delta \sim 1 + R_{ab}\Omega^{a}\Omega^{b}/6$ behavior, whereas the singularity limit expansion depends on the specific choice of $a(t)$. The higher order derivatives of $\Delta(t,T)$ such as $\Box \Delta^{1/2},\Box\Box \Delta^{1/2}...$ contribute crucially to Seeley-deWitt coefficients in a heat kernel expansion and are often expressed in terms of geometric invariants in coincidence limits, such as \cite{Alvarez:2022hjn},
\begin{eqnarray}
    [\Box \Delta^{1/2}] &=& R/6 \\
    \left[ \Box\Box \Delta^{1/2} \right] &=& \frac{1}{5}\Box R + \frac{1}{36}R^{2}-\frac{1}{30}R_{ab}^{2} - \frac{1}{30}R_{abcd}^{2}
\end{eqnarray}

The singularity and coincidence limits of $\Box \Delta^{1/2}$ turns out to be computationally tractable in our case and are given by,

\begin{enumerate}
\item Matter-dominated FLRW spacetime
\begin{eqnarray}\label{eq:BoxDelta_coincidence}
  \textrm{Coincidence}: \ \  \ \Box\Delta^{1/2}(t,T) &=& \frac{2}{9T^{2}}  \\
   \textrm{Singularity}: \ \ \  \Box\Delta^{1/2}(t,T) &\simeq& \frac{2267}{2187\sqrt{21}t^{2}}
\end{eqnarray}

\item Radiation-dominated FLRW spacetime
\begin{eqnarray}\label{eq:BoxDelta_singularity}
  \textrm{Coincidence \& Singularity}: \ \ \  \Box\Delta^{1/2}(t,T) &=&0 
\end{eqnarray}
\end{enumerate}

The coincidence limit of $\Box \Delta^{1/2}$ for matter and radiation dominated FLRW agrees with $R/6$ in the regular region of spacetime i.e. $T\neq 0$. However, the coincidence limit of $\Box \Delta^{1/2}$ diverges as we approach the singularity, whereas the singularity limit ($T\to 0$) remains finite. This signals a breakdown of coincidence limit expansion, and asymptotic expansion around the singularity becomes more meaningful. The non-equivalence of coincidence and singularity limits appearing in our analysis can be summarised concisely as order of limits expression:

\begin{eqnarray}
    \lim_{T\to 0} \lim_{t\to T} [t^2 \Box \Delta^{1/2}] = \frac{2267}{2187\sqrt{21}} \neq \lim_{t\to 0}\lim_{T \to 0} [T^2 \Box \Delta^{1/2}] = \frac{2}{9}
\end{eqnarray}

It is worth noting that this behavior is not universal. In a radiation-dominated FLRW spacetime $(R=0)$, $\Box \Delta^{1/2}$ vanishes in both limits, Eq.\ref{eq:BoxDelta_singularity}, and no such order-of-limits ambiguity arises.

\subsection{Bianchi Type I singularities}

In this section, we use method similar to the FLRW case to generalize our results to the world function for the Bianchi type I metric, which is an exact vacuum solution to Einstein's equations representing a spatially homogeneous (independent of $x,y,z$) but an anisotropic universe with different rates of expansion/contraction in different spatial directions. Bianchi type I solutions have been used extensively in studying cosmological singularities, BKL oscillations, and gravitational chaos near singularities \cite{Misner:1969hg}\cite{Belinsky:1970ew}\cite{Belinsky:1982pk}. A Bianchi type I metric is defined as follows,

\begin{eqnarray}\label{eq:metric_bianchi_I}
    ds^{2} = -dt^{2} + a^{2}_{1}(t)dx^{2} + a_{2}^{2}(t)dy^{2} + a_{3}^{2}(t)dz^{2}
\end{eqnarray}

where the $a_{1},a_{2},a_{3}$ are scale factors providing a straightforward anisotropic generalization to FLRW spacetime. It turns out that this anisotropy can be handled analytically using three spacelike killing vectors corresponding to translations in $(x,y,z)$, giving us three integration constants to deal with. Moreover, we only work in a normal convex neighborhood  $\mathcal{N}$ where a timelike geodesic between two points is unique with time scale $\epsilon=t-T$ restricted by the initial expansion of the geodesics: $\epsilon < 3/|\Theta_{0}|$ as described in the previous section. The world function for this case is defined as follows,

\begin{eqnarray}\label{eq:sigma_bianchi_def}
    \Omega(t,T;\alpha_{1},\alpha_{2},\alpha_{3}) = -\frac{\tau^{2}}{2} = - \frac{1}{2}\left[ \int_{T}^{t}\left( 1 - \sum_{i=1}^{3} \frac{\alpha_{i}^{2}}{a_{i}^{2}(t)+\alpha_{i}^{2}}\right)^{1/2} dt \right]^{2}
\end{eqnarray}

where $\alpha_{1},\alpha_{2},\alpha_{3}$ are integration constants. The Eq.\eqref{eq:sigma_bianchi_def} can be inverted using \emph{Lagrange-Good} formula \cite{Abdesselam:2002mj} for multivariables to give a series expansion in $(x-X),(y-Y),(z-Z)$, where

\begin{eqnarray}
        (x_{i}-X_{i})= \int_{T}^{t}\frac{\alpha_{i}}{a_{1}(t')(a_{1}^{2}(t')+ \alpha_{i}^{2})^{1/2}}dt'=\alpha_{i} \ \phi_{i}(\alpha_{i})
\end{eqnarray}

to give,

\begin{eqnarray}\label{eq:sigma_bianchi_series}
   \Omega &=& \sum_{i,j,k=0}^{\infty}\mathcal{C}_{ijk}(x-X)^{i}(y-Y)^{j}(z-Z)^{k}\\
    \textrm{where} \ \ \ \mathcal{C}_{ijk} &=& \frac{1}{i!j!k!}\left[\frac{\partial^{i+j+k}}{\partial \alpha_{1}^{i}\partial \alpha_{2}^{j}\partial \alpha_{3}^{k}}\left(\Omega(\alpha_{1},\alpha_{2},\alpha_{3})\phi_{1}^{-i}\phi_{2}^{-j}\phi_{3}^{-k} \det\left(\delta_{ab} + \alpha_{b}\frac{\partial \log\phi_{a}}{\partial \alpha_{b}}\right) \right) \right]_{\alpha_{1},\alpha_{2},\alpha_{3}=0}
\end{eqnarray}

where $\alpha=(\alpha_{1},\alpha_{2},\alpha_{3})$ and $a,b=1,2,3$. This series expansion is analytic as long as $\det{\mathbb{J}}\neq0$.

\begin{eqnarray}
    \det(\mathbb{J})= \det\left(\delta_{ab} + \alpha_{b}\frac{\partial \log\phi_{a}}{\partial \alpha_{b}} \right)
\end{eqnarray}

which guarantees local invertibility of the map $(\alpha_{1},\alpha_{2},\alpha_{3}) \to (x_{i}-X_{i})$. A similar branch cut analysis in complex $(\alpha_1,\alpha_2,\alpha_3)$ space gives the upper bound on $(x_j-X_j).$ Namely, $(x_j - X_j) < |i a_j(t')\phi_j(i a_j(t'))|$, where $t' \in \{T, t\}$ is chosen such that $a_j(t') = \min(a_j(T), a_j(t))$. The $\mathcal{O}((x_{i}-X_{i})^{4})$ expansion for $\Omega$ using Eq.\eqref{eq:sigma_bianchi_series} gives,

\begin{eqnarray}
\Omega &=& -\frac{(t-T)^{2}}{2} + \frac{(t-T)}{2}\sum_{i=1}^{3}I_{2}(a_{i})(x_{i}-X_{i})^{2} + \sum_{i,j=1, i\neq j}^{3} \ \frac{ (t-T)I_{2}(a_{i}a_{j})-I_{2}(a_{i})I_{2}(a_{j}) }{(2I_{2}(a_{i})I_{2}(a_{j}))^{2}}(x_{i}-X_{i})^{2}(x_{j}-X_{j})^{2}\\
   \ \ \ \ &+& \ \ \sum_{i=1}^{3}\frac{(t-T)I_{4}(a_{i}) - (I_{2}(a_{i}))^{2}}{8(I_{2}(a_{i}))^{4}}(x_{i}-X_{i})^{4} + \mathcal{O}\left((x_{i}-X_{i})^{6}\right) 
\end{eqnarray}

where,

\begin{eqnarray}\label{eq:I_def}
    I_{n}(a_{i}) := \int_{T}^{t}\frac{dt'}{a(t')^{n}} 
\end{eqnarray}
Note that this series reduces to the isotropic FLRW case of the earlier section when $a_{1}=a_{2}=a_{3}=a$ and reduces further to the Minkowski limit when $a=\textrm{constant}$. A potentially interesting case in Bianchi I models for theoretical interest is the Kasner limit of the Schwarzschild \cite{Mcmaken:2021isj}. In the vicinity of the Schwarzschild singularity, the metric can be brought into a Kasner form with exponents $(2/3,2/3,-1/3)$. The detailed derivation is given in Appendix~\ref{app:kasner_schwarzschild}.

\begin{eqnarray}\label{eq:kasner_scale_factor}
        a_{1}(t) &=& (t/t_{0})^{2/3} \\
        a_{2}(t) &=& (t/t_{0})^{2/3} \\
        a_{3}(t) &=& (t/t_{0})^{-1/3} 
\end{eqnarray}

with $t_{0}=\sqrt{2M}$. The world function to $\mathcal{O}\left((x_{i}-X_{i})^{2}\right)$ for this case is given by,
\begin{eqnarray}\label{eq:sigma_schw}
    \Omega_{\textrm{schw}}(t,T;\ell_{\perp},z-Z) &=& -\frac{(t-T)^{2}}{2} + \frac{(t-T)}{6t_{0}((t_{0}/T)^{1/3} - (t_{0}/t)^{1/3})}  \ell^{2}_{\perp}
    \\
       \ \  \ \ \ \ \ \   &+& \frac{5(t-T)}{6(t(t/t_{0})^{2/3} - T(T/t_{0})^{2/3})} (z-Z)^{2} + \mathcal{O}\left((x_{i}-X_{i})^{4}\right)
\end{eqnarray}

Where $\ell^{2}_{\perp}= (x-X)^{2}+(y-Y)^{2}$. It is easy to verify that Eq.\eqref{eq:sigma_schw} satisfies the necessary coincidence limits of a world function i.e. $[\Omega_{\rm schw}]=[\nabla_{a}\Omega_{\rm schw}]=0\ , \ [\nabla_{a}\nabla_{b}\Omega_{\rm schw}]=g_{ab}=\text{diag}(-1,(t/t_{0})^{2/3},(t/t_{0})^{2/3},(t/t_{0})^{-1/3})$ and gives a self consistent series expansion of $\Omega$ in Schwarzschild case. However, this series expansion is strictly valid only within the domain of analyticity, the maximum spatial separations in the transverse (expanding) and longitudinal (contracting) directions are constrained by $\ell_\perp < L_{\perp,\text{max}}$ and $|z - Z| < L_{z,\text{max}}$, respectively, where:

\begin{align}
L_{\perp,\text{max}} &= 3 t_0^{2/3} T^{1/3} \left[ \frac{\sqrt{\pi} \Gamma(5/4)}{\Gamma(3/4)} - \left(\frac{T}{t}\right)^{1/3} {}_2F_1\left(\frac{1}{4}, \frac{1}{2}, \frac{5}{4}; \left(\frac{T}{t}\right)^{4/3}\right) \right], \\[1em]
L_{z,\text{max}} &= \frac{t_0^{-1/3}}{8} \left[ 3 \left(3 t^{2/3} + 2 T^{2/3}\right) (t T)^{1/3} \sqrt{1 - \left(\frac{T}{t}\right)^{2/3}} + 9 t^{4/3} \operatorname{arcsec}\left(\left(\frac{t}{T}\right)^{1/3}\right) \right].
\end{align}

Furthermore, we calculated the van Vleck determinant for the Kasner limit of Schwarzschild spacetime, which shows different scaling behavior in \emph{singular} and \emph{coincidence} limits.

\begin{eqnarray}\label{eq:Delta_schw}
        \textrm{Coincidence limit} \ \ \ : \ \ \Delta_{\text{schw}} &=& 1 + \mathcal{O}(\varepsilon^{4})  \ \\
         \textrm{Singularity limit} \ \ \ : \ \  \Delta_{\text{schw}} &=& -\frac{5\ell_{\perp}^{2}(t/t_{0}^{4})^{1/3}}{486\ T}   -\frac{10\ell_{\perp}^{2}}{243\ T^{2/3} \ t_{0}^{4/3}} + \frac{25t_{0}^{2/3}(z-Z)^{2}}{286\ t^{7/3} T^{1/3} \ t_{0}^{4/3}} \\&& \ \ - \frac{25\ell_{\perp}^2}{243\ (t \ t_{0}^4)^{1/3}T^{1/3}} + \frac{5t^{1/3}}{27 T^{1/3}} + \frac{10}{27} + ... 
\end{eqnarray}

The first non-trivial contribution in coincidence limit expansion of $\Delta_{\textrm{schw}}$ comes at order $\mathcal{O}(\varepsilon^{4})$ because order $\mathcal{O}(\varepsilon^{2})$ is identically zero for Schwarzschild spacetime ($R_{ab}=0$). The singularity limit of $\Delta_{\textrm{schw}}$ shows a $1/T^{1/3}$ divergence (fifth term in Eq.\eqref{eq:Delta_schw}) in $\ell$-independent terms as compared to $1/T$ divergence (third term in Eq.\eqref{eq:Delta_FLRW}) observed in the case of matter-dominated FLRW spacetime; the different scaling behaviour can be attributed to the presence of non-zero shear in Kasner (Schwarzschild) spacetime as compared to an isotropic expansion in matter-dominated FLRW spacetime. Again, the leading divergent behaviour in the singularity limit depends on the $\ell_{\perp}$ and ($z-Z$) independent terms since $\ell_{\perp} \lesssim t_{0}^{2/3}T^{1/3}$ and $(z-Z) \lesssim t^{4/3}t_0^{-1/3}$.

%%%%%%%%%%%%%%%%%%%%%%%%%%%%%%%%%%%%%%%%%%%%%%%%%%%%%%%%%%%%%%%%%%

%%%%%%%%%%%%%%%%%%%%%%%%%%%%%%%%%%%%%%%%%%%%%%%%%%%%%%%%%%%%%%%%%%%%%%%%
\section{Geodesic and causal structure near singularity}

\begin{figure}[htbp]\label{fig:relative_expansion_Omega}
    \centering
    \includegraphics[width=1\linewidth]{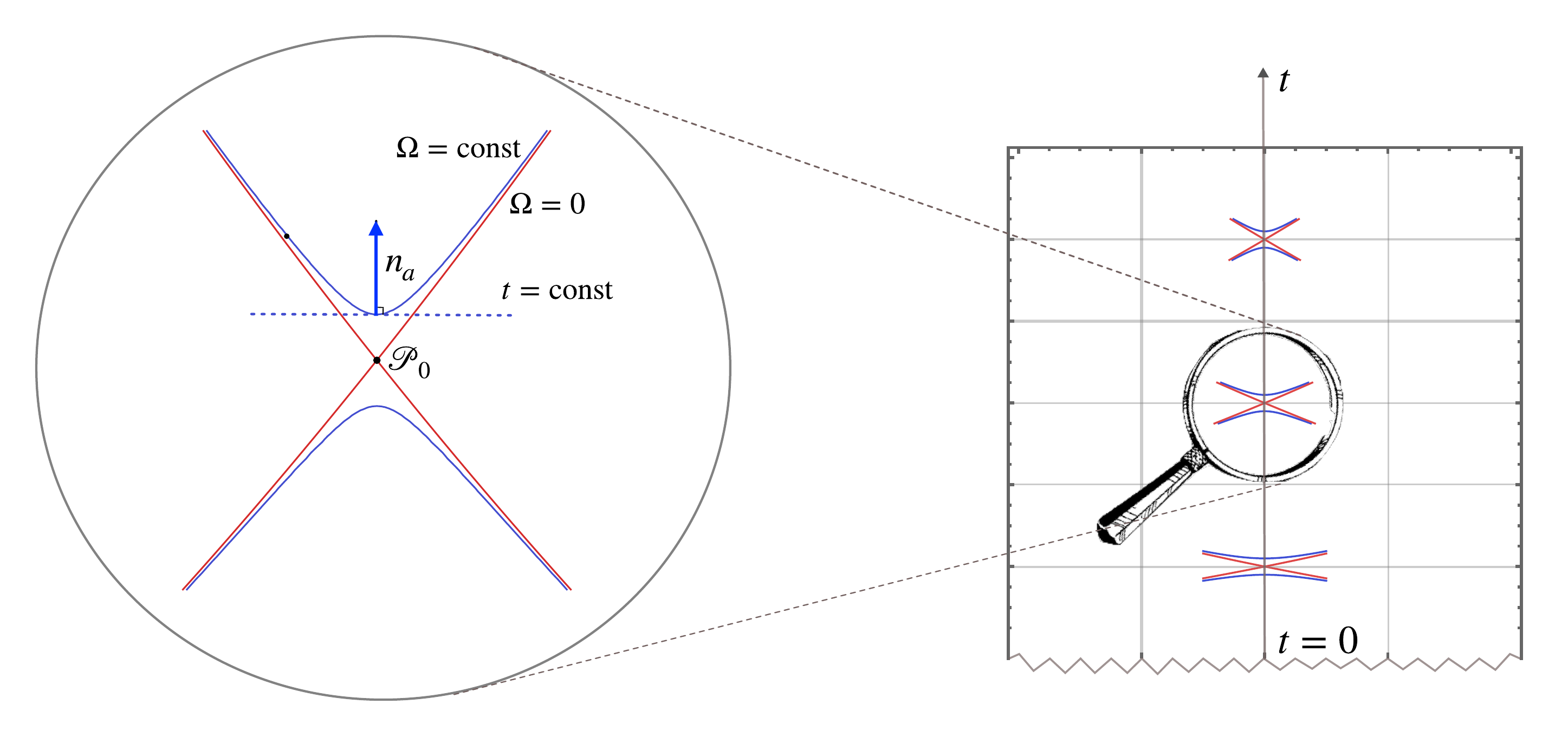}
    \caption{A local region of matter-dominated FLRW spacetime is shown with $t=$constant and $\Omega=$ constant anchored at a point $P(T, X, Y, Z)$. The extrinsic curvature $K_{ab}$ of $\Omega$=constant hypersurfaces characterizes the local expansion of these surfaces about point $P$. The $t=$constant hypersurfaces follow Hubble's law, whereas $\Omega=$constant hypersurfaces obey a generalized version of Hubble's law \cite{Kothawala:2018BGV}.}
\end{figure}

The equi-geodesic surfaces are defined as surfaces $\Omega(p,\mathscr{P})=$ const. about a spacetime point $\mathscr{P}$. These surfaces constitute a fundamental geometric structure which has been extensively explored in \cite{Stargen:2015hwa}\cite{Kothawala:2013maa}. We will now consider the behaviour of these equi-geodesic surface $\Sigma(\kappa)$ defined by $\{p \  | \  \Omega(p,P)=-\kappa \ ,\ p \in \mathcal{M}\}$ in a FLRW spacetime with one of the points of world function $\Omega$ anchored at $P=(T, X, Y, Z)$. The equations for $\Sigma(\kappa)$ and lightcones are given by, 

\begin{eqnarray}\label{eq:Sigma_def_eq:lightcone_def}
    \Omega &=& -\kappa = -\frac{(t-T)^{2}}{2} + \frac{(t-T)}{2I_{2}}\ell^{2}\\
     \ell &=&\pm\int_{T}^{t}\frac{dt'}{a(t')} 
\end{eqnarray}

\begin{figure}[htbp]
    \centering
    \includegraphics[width=.46\linewidth]{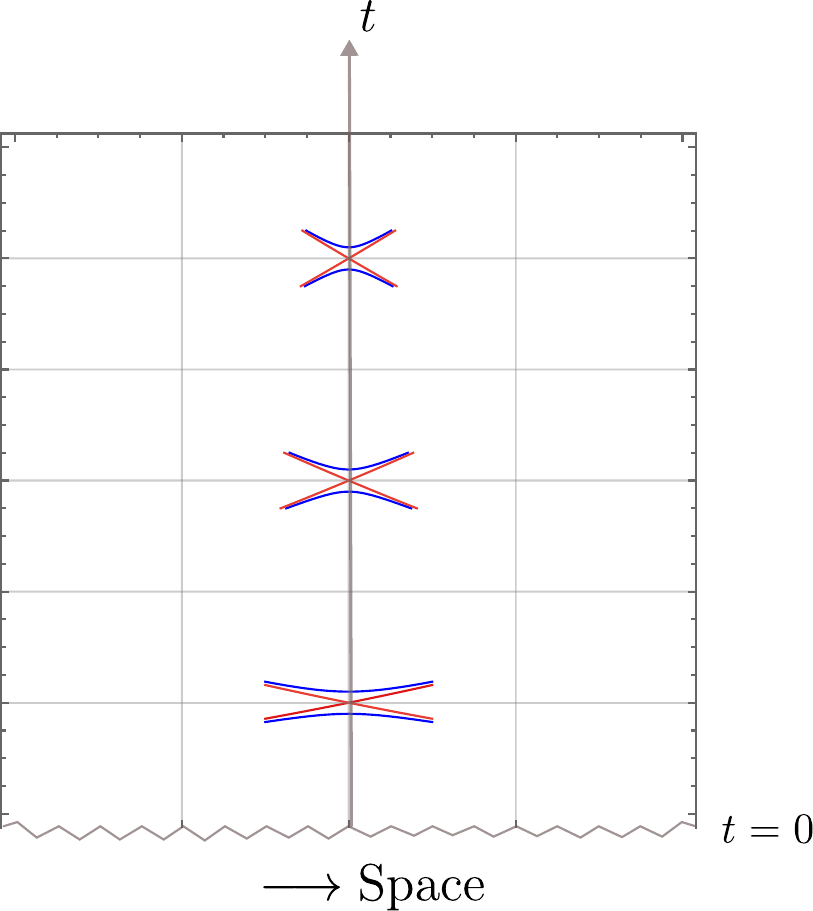}
    \hskip 12pt
    \centering
    \includegraphics[width=.46\linewidth]{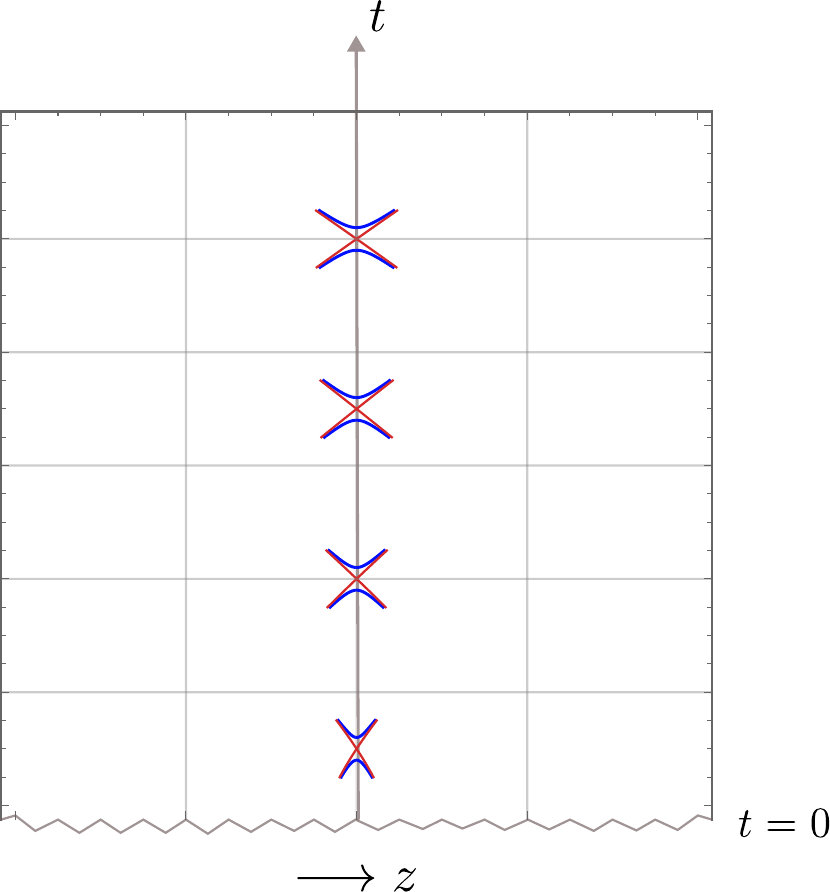}
  \caption{The equi-geodesic surfaces and lightcones are constructed using biscalar $\Omega(x^{a}, X^{a})$ about different points $P$ in the spacetime. \textbf{Left}: Matter-dominated FLRW spacetime with $p=2/3$, \textbf{Right}: Kasner form of Schwarzschild singularity. The zig-zag line shows the curvature singularity in both cases.}
\label{fig:equi-geodesic_surface}
\end{figure}

We plot the set of Eqs. \eqref{eq:Sigma_def_eq:lightcone_def} in the vicinity of the singularity $\mathscr{P}$ about different points $P$ translated in time to study the causal structure near the singularity. Since FLRW spacetime is homogeneous, we consider $X, Y, Z=0$ without loss of generality. The light cones flatten out in the FLRW case as seen from Fig. \ref{fig:equi-geodesic_surface} (left). We find a contrasting behavior for Schwarzschild geometry, as is evident from Fig. \ref{fig:equi-geodesic_surface} (right), which shows light cones drawn at different points $P$ along the radial infalling geodesic of a point particle in the vicinity of the Schwarzschild singularity begin to shrink and converge as we approach the singularity along the \emph{contracting} dimension.\\

To complement this causal characterization of spacetime, we now turn to a geometric description based on equi-geodesic surfaces, whose asymptotic structure is governed by the light cones. The equi-geodesic surfaces can be locally characterized by their extrinsic curvature $K$ as demonstrated by one of the authors in \cite{Stargen:2015hwa}. We use the $\mathcal{O}(\ell^{2})$ expression for the world function of the FLRW metric to derive an explicit expression for the extrinsic curvature $K_{ab}$ for $\Omega=\text{constant}$ surfaces in this spacetime. The normal $n_{a}$ to the $\Omega=-\kappa=\textrm{constant}$ surface is defined as,

\begin{eqnarray}
    n_{a} = \frac{\nabla_{a}\Omega}{2\sqrt{-\Omega}}
\end{eqnarray}

The extrinsic curvature $K_{ab}$ and trace $K$ for this normal in FLRW evaluates to,

\begin{eqnarray}\label{eq:K_ab}
    K_{ab}=\nabla_{a}n_{b} = \left(\frac{1}{(t-T)F(t,T)}+a(t)\dot{a}(t)\right)\begin{pmatrix} 0 &0 & 0 & 0\\ 0& 1 & 0 & 0 \\ 0 & 0 & 1 & 0 \\ 0 & 0 & 0 & 1
    \end{pmatrix}
\end{eqnarray}

\begin{eqnarray}\label{eq:K_trace}
    K = \textrm{Tr}(K_{ab})=\frac{3}{(t-T)F(t,T)a^{2}(t)} + \frac{3\dot{a}(t)}{a(t)}
\end{eqnarray}

The trace $K=K(t,T)$ as defined above is a biscalar. The coincidence limit expansion of Eq.\eqref{eq:K_trace} in expansion parameter $\varepsilon = t-T$ gives,

\begin{eqnarray}\label{eq:K_coincidence}
%    K \simeq -\frac{3}{\varepsilon} + \frac{3 \dot{a}(t)}{a(t)} - \frac{3 \dot{a}(t)}{a(t)} + \mathcal{O}(\varepsilon) = -\frac{3}{\varepsilon} + \mathcal{O}(\varepsilon)
%
  \textrm{Coincidence}: \ \ \ \   K = \frac{3}{\varepsilon} +\frac{\ddot{a}(t)}{a(t)}\varepsilon + \mathcal{O}(\varepsilon^2)
\end{eqnarray}

The $\varepsilon^{-1}$ divergence term is a universal divergence that is independent of the scale factor $a(t)$ and can be attributed to the flat space limit. The second term in Eq.\eqref{eq:K_coincidence} corresponds to the tidal trace and matches with another result obtained by one of the author in a more general setting \cite{Stargen:2015hwa}. The universal divergent term in $K$ can be renormalized by subtracting out $K_{0}=3/\varepsilon$ to give,

\begin{eqnarray}\label{eq:K_ren}
    \tilde{K} = K-K_{0} = \frac{\ddot{a}(t)}{a(t)}\varepsilon + \mathcal{O}(\varepsilon^{2})
\end{eqnarray}

Thus, if we have a local construction of $\Omega=\textrm{constant}$ surface about a point $(T, X, Y, Z)$ in its convex neighbourhood, the extrinsic curvature of that surface depends on the second derivative of the scale factor! It is quite interesting to note that this is distinct from the \emph{Hubble} expansion term $3\dot{a}/a$.\\

On the other hand, in the singularity 
limit, $T\to 0,t\to 0^+$, $K$ evaluates to

\begin{eqnarray}
  \textrm{Singularity}: \ \ \ \   K \simeq \frac{3q}{t} 
\end{eqnarray}

The leading term appearing in this expression is nothing but the expansion scalar $\Theta$ for $t=$ constant surfaces defined by 
    $\Theta := d (\ln V)/dt = {3q}/{t}$ 
    with $V \propto a^{3} \propto (t/t_{0})^{3q}$. This peculiar result can be understood from the fact that, in FLRW spacetimes, the $\Omega=\textrm{constant}$ surfaces and $t=\textrm{constant}$ surfaces tend to merge smoothly as we approach the singularity; this is illustrated by explicit plots of these surfaces for matter-dominated case given in Fig.\ref{fig:equi-geodesic_surface}. \\
    
\begin{figure}[htbp]
    \centering
    \includegraphics[width=0.46\linewidth]{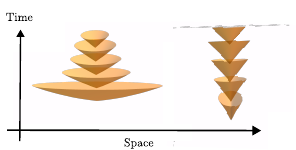}
     \hskip 1pt
    \centering
    \includegraphics[width=.48\linewidth]{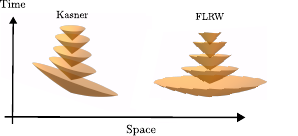}
    \caption{The lightcones near a Kasner (Schwarzschild) singularity stretch due to non-zero shear and anisotropic nature of spacetime, whereas lightcones in FLRW spacetime expand isotropically. \textbf{Left}: Lightcones in Kasner (Schwarzschild) elongate (left) along the expanding direction and squeeze (right) along the contracting direction (here $z$). \textbf{Right}: General comparison of lightcone structure between Kasner (Schwarzschild) and FLRW (matter-dominated) spacetime.}
    \label{fig:3Dkasner}
\end{figure}

    The extrinsic curvature and its trace for Kasner (Schwarzschild) evaluates to,

    \begin{eqnarray} 
         K_{ab}=\nabla_{a}n_{b} &=& \sum_{a,b=1}^{3}\left(\frac{1}{I_{2}(a_{i})}+a_{i}(t)\dot{a}_{i}(t)\right)\delta_{ab}\\
      \textrm{Singularity}:   K &=& \frac{8}{3t}\\
      \textrm{Coincidence}:  K &=&  \frac{3}{\varepsilon} - \frac{8\varepsilon^{3}}{1215T^{4}}+....
    \end{eqnarray}

    where $a_{i}$ are defined as in Eq.\eqref{eq:kasner_scale_factor}. The behaviour of lightcones in this case should be contrasted with the Kasner case for Schwarzschild singularity (see Fig. \ref{fig:3Dkasner} and right half of Fig. \ref{fig:equi-geodesic_surface}) where lightcones close up along the contracting dimension (here $z$), reducing the region of spacetime that remains in causal contact for the timelike observer as singularity is approached.

\section{Discussion}\label{sec:concluding-sec}

In this work, we derive a local series representation for Synge’s world function in FLRW and Bianchi type~I spacetimes, valid within a normal convex neighbourhood $\mathcal{N}$ with $\varepsilon=t-T < 3/|\Theta_0|$ and $\ell< L_{\textrm{max}}$ to avoid caustics and ensure convergence of series. Specifically in FLRW, the closed form representation for $\Omega(t,T;\ell)$ is an analytic function of $t,T,\ell$ as long as scale factor is non-zero along the geodesic connecting the two points and $\ell$ is bounded by the minimum of scale factor at the endpoints, $\textrm{min}(a(T),a(t))$. Interestingly, the key structural properties of the world function are captured through integrals of $1/a(t')^{2n}$, $n \in \mathbb{N}$, between the end-points of a given timelike geodesic provided $a(t')\neq0$ for $t' \in (T,t)$. Such dependence was also observed by Buchdahl long back \cite{Buchdahl:1972rw} in the expansion of $\sqrt{\Omega}$, but this expansion turns out to be bad near singularities due to spurious, unphysical divergences. Our representation, on the other hand, gives well-defined limits if one of the points is singular.

As an important concomitant, we have computed the singular and coincidence limits of the van Vleck determinant for FLRW (matter-dominated and radiation) and Bianchi class of spacetimes, focusing in particular on the Kasner limit of the Schwarzshild near the singularity. The van Vleck determinant shows a divergent power law behaviour in the singular limit for matter-dominated FLRW and Kasner cases, whereas a logarithmic divergence for the radiation-dominated FLRW case. We provide some further insights into the geodesic structure close to singularities by investigating the behavior of lightcones and extrinsic curvature for the equi-geodesic ($\Omega=$ constant) surfaces anchored at a point, which are asymptotic to these lightcones, thereby characterizing the causal structure of the spacetime. The trace of extrinsic curvature gives us the expansion of $\Omega=$ constant surfaces, and we compare this with the expansion of $t=$constant surface; the differences are interesting and shown in 
Fig. \ref{fig:relative_expansion_Omega}. 

The contrasting behaviour of coincidence and singularity limits of biscalars suggests that the spacetime neighbourhood near a singularity is qualitatively different from a regular region of spacetime. Given the fundamental significance of world function and van Vleck determinant, this fact has important implications. We list a few below:
\begin{enumerate}
    \item Point splitting regularization and structure of two point functions near singularities. 
    \item Study of WKB wave function of quantum systems near singularities.
    \item Causal structure near generic singularities.
    \item Schwinger-de Witt expansion for the heat kernel near singularities.
    \item Small scale structure of quantum spacetime. Several recent works have demonstrated the important role of bi-tensors in characterizing the small-scale structure of spacetime; \cite{Kothawala:2013maa},\cite{Kothawala:2023tuh},\cite{qmetric-rev1}, and our work provides important tools to study this structure near the singularity, thereby providing a window into quantum effects near the singularity.
    \end{enumerate}

%%%%%%%%%%%%%%%%%%%%%%%%%%%%%%%%%%%%%%%%%%%%%%%%%%%%%%%%%%%%%%%%%%%%%%%%

\appendix
\label{FLRW_omega_derivation}
\section{Calculational details in FLRW spacetime}

We can reduce the solution of calculating proper time to quadrature (see \cite{Roberts:1993umg}) and get the following expressions,

\begin{eqnarray}\label{eq:G_def}
    \tau(t,T;\alpha) &=& \int_{T}^{t} \frac{a(t')}{\sqrt{\alpha^{2} + a^{2}(t')}}dt'\\
    \ell(t,T;\alpha) &=& {\alpha} \int_{T}^{t} \frac{dt'}{a(t')\sqrt{\alpha^{2} + a^{2}(t')}}= \alpha G(\alpha)
\end{eqnarray}

where $\alpha$ is an integration cosntant, Now $\Omega$ can be rewritten as (using $d\tau^{2}/dt' = 2\tau (d\tau/dt')$),

\begin{eqnarray}
    \Omega(t,T;\alpha) &=& -\frac{\tau^{2}}{2} =-\int_{T}^{t} \left(\tau \frac{d\tau}{dt'}\right) dt'\\
     &=& - \int_{T}^{t} \left[\int_{T}^{t'} \frac{a(t_{1})}{\sqrt{a^{2}(t_{1})+\alpha^{2}}}dt_{1}\right]\frac{a(t')}{\sqrt{\alpha^{2}+ a^{2}(t')}}dt' 
\end{eqnarray}

So now we have $\Omega(\alpha)$ to be expanded using the \emph{Lagrange-Good} formula \cite{Abdesselam:2002mj} in powers of $\ell$. Thus expansion in $\ell$ given as,

\begin{eqnarray}
    \Omega(t,T;\ell) := \Omega(0) + \sum_{k=1}^{\infty}\frac{\ell^{k}}{k!}\left[ \frac{d^{k-1}}{d\alpha^{k-1}}({\Omega}(\alpha)' G^{-k}(\alpha))
    \right]_{\alpha=0}
\end{eqnarray}

where,

\begin{eqnarray}\label{eq:F_def}
    \frac{d\Omega}{d\alpha}= \alpha \int_{T}^{t}\int_{T}^{t'}\frac{dt_{1}dt'a(t_{1})a(t')}{\sqrt{\alpha^{2}+ a(t_{1})^{2}}\sqrt{\alpha^{2}+ a(t')^{2}}}\left[ \frac{1}{\alpha^{2}+a(t')^{2}} + \frac{1}{\alpha^{2}+a(t_{1})^{2}} \right] = \alpha F(\alpha)
\end{eqnarray}

It is easy to confirm that $\Omega(0)=-(t-T)^{2}/2$, which makes sense in $\ell \longrightarrow 0$ limit. Since functions $\Omega, G, F$  are all even functions of $\alpha$, only even terms contribute in the expansion.

\begin{eqnarray}
    \Omega(t,T;\ell)=-\frac{(t-T)^{2}}{2} + \sum_{j=1}^{\infty}\frac{\ell^{2j}}{(2j)!}\left[\frac{d^{2j-1}}{d\alpha^{2j-1}} (\alpha F)(G)^{-2j} \right]_{\alpha = 0}
\end{eqnarray}

where $F$ and $G$ are defined in Eqs. \eqref{eq:F_def} and \eqref{eq:G_def}. Let us now consider $j=1$ term explicitly, The expansion for $G$ is,

\begin{eqnarray}
    G(\alpha) = \sum_{n=0}^{\infty}\left( \nobarfrac{-1/2}{n}\right) I_{2n+2}\alpha^{2n}\simeq I_{2} - \frac{1}{2}I_{4}\alpha^{2} + \mathcal{O}(\alpha^{4})
\end{eqnarray}

where,

\begin{eqnarray}
    I_{n}(a) = \int_{T}^{t}\frac{dt_{1}}{a(t_{1})^{n}}
\end{eqnarray}

Let us now consider expansion for $F(\alpha)$,

\begin{eqnarray}
    I = \int_{T}^{t}\int_{T}^{t'}\frac{dt_{1}dt' a(t_{1})a(t')}{(\alpha^{2}+ a(t_{1})^{2})^{1/2}(\alpha^{2}+ a(t')^{2})^{3/2}} =  \int_{T}^{t} \left[\int_{T}^{t'} \frac{a(t_{1})dt_{1}}{(\alpha^{2}+ a^{2}(t_{1}))^{1/2}}\right] \frac{dt'a(t')}{(\alpha^{2} + a^{2}(t'))^{3/2}}dt'
\end{eqnarray}

Now, the integral in the square bracket can be expanded in a series as,

\begin{eqnarray}
    \int_{T}^{t'} \frac{a(t_{1})dt_{1}}{(\alpha^{2}+ a^{2}(t_{1}))^{1/2}} = \sum_{k=0}^{\infty}\left( \nobarfrac{-1/2}{k}\right) I_{2k}(t')\alpha^{2k}
\end{eqnarray}

Thus,

\begin{eqnarray}
    \int_{T}^{t} \left[\sum_{k=0}^{\infty}\left( \nobarfrac{-1/2}{k}\right) I_{2k}(t')\alpha^{2k} \right]\frac{dt' a(t')}{(\alpha^{2} + a^{2}(t'))^{3/2}} = \int_{T}^{t} \left[\sum_{k=0}^{\infty}\left( \nobarfrac{-1/2}{k}\right) I_{2k}(t')\alpha^{2k}\right]\sum_{m=0}^{\infty}\left( \nobarfrac{-3/2}{m}\right)\frac{\alpha^{2m}}{a^{2m+2}(t')}dt'
\end{eqnarray}

Therefore expression for $F$ is given by,

\begin{eqnarray}
    F = \sum_{k,m=0}^{\infty}\alpha^{2(k+m)}\left( \nobarfrac{-1/2}{k} \right)\left( \nobarfrac{3/2}{m} \right) \int_{T}^{t} \frac{I_{2k}(t')}{a^{2m+2}(t')}dt' +  \ \  (t_{1} \leftrightarrow t')
\end{eqnarray}

where $ (t_{1} \leftrightarrow t')$ is given by,

\begin{eqnarray}
     (t_{1} \leftrightarrow t') &=&  \int_{T}^{t} \left[\int_{T}^{t'} \frac{a(t_{1})dt_{1}}{(\alpha^{2}+ a^{2}(t_{1}))^{3/2}}\right] \frac{dt'a(t')}{(\alpha^{2} + a^{2}(t'))^{1/2}}dt'\\ &=& \int_{T}^{t} \sum_{k=0}^{\infty}\left( \nobarfrac{-3/2}{k} \right)\alpha^{2k}\int_{T}^{t'}\frac{dt_{1}}{a^{2k+2}(t_{1})}\sum_{m=0}^{\infty}\left( \nobarfrac{-1/2}{m} \right)\frac{\alpha^{2m}}{a^{2m}(t')}dt'\\
     &=& \sum_{k,m=0}^{\infty}\alpha^{2(k+m)}\left( \nobarfrac{-3/2}{k} \right)\left( \nobarfrac{-1/2}{m} \right)\int_{T}^{t}\frac{I_{2k+2}(t')}{a^{2m}(t')}dt'
\end{eqnarray}

Therefore, the complete expression for $F$ is given by,

\begin{eqnarray}
    F &=& \sum_{k,m=0}^{\infty}\alpha^{2(k+m)}\left( \nobarfrac{-1/2}{k} \right)\left( \nobarfrac{-3/2}{m} \right) \int_{T}^{t} \frac{I_{2k}(t')}{a^{2m+2}(t')}dt' + \sum_{k,m=0}^{\infty}\alpha^{2(k+m)}\left( \nobarfrac{-3/2}{k} \right)\left( \nobarfrac{-1/2}{m} \right)\int_{T}^{t}\frac{I_{2k+2}(t')}{a^{2m}(t')}dt'\\
      &=& \sum_{k,m=0}^{\infty}\alpha^{2(k+m)}\left( \nobarfrac{-1/2}{k} \right)\left( \nobarfrac{-3/2}{m} \right)\left[\int_{T}^{t} \frac{I_{2k}(t')}{a^{2(m+1)}(t')}dt' + \int_{T}^{t} \frac{I_{2m+2}(t')}{a^{2k}(t')}dt'  \right]\\
      &=& \sum_{k,m=0}^{\infty}\alpha^{2(k+m)}\left( \nobarfrac{-1/2}{k} \right)\left( \nobarfrac{-3/2}{m} \right)\int_{T}^{t}\left[\frac{I_{2k}(t')}{a^{2(m+1)}(t')} + \frac{I_{2m+2}(t')}{a^{2k}(t')} \right]dt'
\end{eqnarray}

Let's calculate lowest order contribution in $\alpha$ to $(-\alpha F (G^{-1})^{2})$, i.e. $n=k=m=0$ in series for $G$ and $F$.\\ Therefore,

\begin{eqnarray}
        G_{0} &=& I_{2}(t) = \int_{T}^{t}\frac{dt_{1}}{a^{2}(t_{1})}\\
        F_{0} &=& F|_{m,k=0} = \int_{T}^{t}\left[ \frac{I_{0}(t')}{a^{2}(t')}+ I_{2}(t')\right]dt'\\
        &=& \int_{T}^{t}\left[ \frac{t'-T}{a^{2}(t')} + \int_{T}^{t'}\frac{dt_{1}}{a^{2}(t_{1})} \right]dt'\\
        &=& (t'-T)|_{T}^{t} \int_{T}^{t}\frac{dt'}{a^{2}(t')}- \int_{T}^{t}\int_{T}^{t'}\frac{dt_{1}}{a^{2}(t_{1})}dt' + \int_{T}^{t}\int_{T}^{t'}\frac{dt_{1}}{a^{2}(t_{1})}dt' \\
        &=&(t-T)\int_{T}^{t}\frac{dt'}{a^{2}(t')}dt'
\end{eqnarray}

Therefore $\mathcal{O}(\ell^{2})$ contribution to $\Omega$ is given by,

\begin{eqnarray}
    \Omega(t,T;\ell) = -\frac{(t-T)^{2}}{2}+ \frac{(t-T)}{2}\left[\int_{T}^{t}\frac{dt'}{a^{2}(t')} \right]^{-1}\ell^{2} + \mathcal{O}(\ell^{4})
\end{eqnarray}

Consider the expression,

\begin{eqnarray}
    \Omega(t,T;\ell) &=& \Omega(0)  + \sum_{j=1}^{\infty}\frac{\ell^{2j}}{(2j)!}\left[\frac{d^{2j-1}}{d\alpha^{2j-1}} \Bigl\{ (\alpha F)(G^{-1})^{2j} \Bigr\} \right]_{\alpha = 0}\\
    &=& \Omega(0) + \sum_{j=1}^{\infty}\frac{\ell^{2j}}{(2j)!}\left[\frac{d^{2j-1}}{d\alpha^{2j-1}} \Bigl\{ \frac{d\Omega}{d\alpha}(G^{-1})^{2j} \Bigr\} \right]_{\alpha = 0} 
\end{eqnarray}

Consider the series separately.

\begin{eqnarray}
    \frac{d^{2j-1}}{d\alpha^{2j-1}}\left( \frac{d\Omega}{d\alpha}(G^{-1})^{2j} \right)&=&\frac{d^{2j-1}}{d\alpha^{2j-1}} \left( \frac{d}{d\alpha}(\Omega (G^{-1})^{2j}) - \Omega \ 2j (G^{-1})^{2j-1}(G^{-1})' \right)  \\
    &=& \frac{d^{2j-1}}{d\alpha^{2j-1}}\left(  \frac{d}{d\alpha}(\Omega (G^{-1})^{2j}) + 2j \ \Omega (G^{-1})^{2j-1}(G^{-1})' \right)\\
    &=& \frac{d^{2j}}{d\alpha^{2j}}(\Omega (G^{-1})^{2j}) + 2j \frac{d^{2j-1}}{d\alpha^{2j-1}}[\Omega (G^{-1})^{2j-1}(G^{-1})']
\end{eqnarray}

Thus,

\begin{eqnarray}
    \Omega(t,T;\ell)-\Omega(0)= \sum_{j=1}^{\infty}\frac{\ell^{2j}}{(2j)!}\frac{d^{2j}}{d\alpha^{2j}}(\Omega(\alpha) (G)^{-2j}) + \ell \sum_{j=1}^{\infty}\frac{\ell^{2j-1}}{(2j-1)!} \frac{d^{2j-1}}{d\alpha^{2j-1}}[ \Omega(\alpha) (G)^{-2j+1}((G)^{-1})']
\end{eqnarray}

Note that we have the following identification,

\begin{eqnarray}
        \cosh(\ell D_{\alpha} [(G)^{-1}...])-1 &:=& \sum_{j=1}^{\infty}\frac{(\ell D_{\alpha})^{2j}[(G)^{-2j}...]}{(2j)!}\\
        \sinh(\ell D_{\alpha}[(G)^{-1}...]) &:=& \sum_{j=1}^{\infty}\frac{(\ell D_{\alpha})^{2j-1}[(G)^{-2j+1}...]}{(2j-1)!}
\end{eqnarray}

so that the final expression for the world function can be written concisely as follows,

\begin{eqnarray}
    \Omega(t,T;\ell) = \Omega(0) + \cosh(\ell D_{\alpha}[G^{-1}\Omega(\alpha)])  + \ell \sinh(\ell D_{\alpha}[G^{-1}(G^{-1})'\Omega(\alpha)])
\end{eqnarray}

\section{Van Vleck Determinant in de Sitter Spacetime}
\label{app:vvd_desitter}

In this section of appendix, we verify the general perturbative expansion of the Van Vleck-Morette determinant derived from the world function ansatz. We achieve this by evaluating the series expansion for a $(3+1)$-dimensional de Sitter spacetime and comparing it against the known exact analytical expression. This comparison serves as a non-trivial consistency check for the coefficients of the world function, particularly the $\mathcal{O}(\ell^4)$ contribution.\\

The world function $\Omega(t,T; \ell)$ for a Friedmann-Lemaître-Robertson-Walker (FLRW) metric, expanded in the spatial coordinate distance squared $\ell^2 = |\mathbf{x} - \mathbf{X}|^2$, is given by

\begin{equation}\label{eq:omega_series_order4}
    \Omega(t,T; \ell) = -\frac{(t-T)^2}{2} + \frac{1}{2F(t,T)}\ell^2 + Q(t,T)\ell^4 + \mathcal{O}(\ell^6),
\end{equation}

where $F(t,T)$ and $Q(t,T)$ are functions of the scale factor $a(t)$ and its integrals.

The Van Vleck determinant $\Delta(x, x')$ is defined as the normalized determinant of the mixed partial derivatives of the world function:

\begin{equation}
    \Delta(x, x') = -\frac{\det \left[ \partial_\mu \partial_{\nu'} \Omega(x, x') \right]}{a(t)^3 a(T)^3}.
\end{equation}

By explicitly calculating the determinant of the $4 \times 4$ Hessian up to $\mathcal{O}(\ell^2)$, we obtain the general series expansion:

\begin{equation}
    \Delta(t, T; \ell) = \Delta_0(t,T) + \Delta_2(t,T) \ell^2 + \mathcal{O}(\ell^4),
\end{equation}

where the coefficients are found to be

\begin{align}
    \Delta_0(t,T) &= \frac{1}{\left[ a(t) a(T) F(t,T) \right]^3}, \\
    \Delta_2(t,T) &= \frac{3(40 F(t,T)^3 Q(t,T) - F^{(1,1)}(t,T))}{2 \left[a(t) a(T)\right]^3 F(t,T)^5},
\end{align}

with $F^{(1,1)}(t,T) \equiv \partial_t \partial_T F(t,T)$. To verify this result, we specialize to de Sitter spacetime, characterized by the scale factor $a(t) = e^{Ht}$. The exact Van Vleck determinant for a $(3+1)$-dimensional de Sitter spacetime is known in closed form \cite{Stargen:2015hwa}:

\begin{equation}
    \Delta_{\text{exact}}(\Omega) = \left( \frac{\sqrt{-\Omega(t,T;\ell) \Lambda}}{\sinh(\sqrt{-\Omega(t,T;\ell) \Lambda})} \right)^3,
    \label{eq:exact_vvd_dS}
\end{equation}

where $\Omega$ is Synge's world function. For two points separated by a coordinate time $\Delta t = t-T$ and a spatial coordinate distance $\ell$ . By substituting Eq.~\eqref{eq:omega_series_order4} into Eq.~\eqref{eq:exact_vvd_dS} and performing a Taylor expansion with respect to $\ell^2$, we obtain an exact analytical series for the de Sitter Van Vleck determinant parameterized by $t,T,\ell$. Simultaneously, we evaluate our general derived coefficients $\Delta_0$ and $\Delta_2$ by substituting $a(t) = a_0e^{H(t-t_0)}$ into the explicit integral definitions of $F(t,T)$ and $Q(t,T)$ (see Eq. \ref{eq:F_and_Q}) to get the following expressions,

\begin{eqnarray}
    \Delta_0(t, T) &=& \left( \frac{H(t-T)}{\sinh[H(t-T)]} \right)^3 \\[10pt]
    \Delta_2(t, T) &=& \frac{3}{2} a(t)a(T) H^4 (t-T)^2 \frac{H(t-T)\coth[H(t-T)] - 1}{\sinh^4[H(t-T)]}
\end{eqnarray}

Comparing the two distinct approaches, we find that the analytical coordinate expansion of $\Delta_{\text{exact}}(s)$ precisely matches the evaluated functions $\Delta_0(t,T)$ and $\Delta_2(t,T)$. This exact agreement at $\mathcal{O}(\ell^2)$ validates the derivation of the van Vleck determinant and confirms the consistency of our world function series representation used in this work.

\section{Kasner form of Schwarzschild singularity}
\label{app:kasner_schwarzschild}

In this section of the appendix, we show that the Schwarzschild interior near the singularity can be written in a Kasner form. The metric near the Schwarzschild black hole of mass $M$ near the singularity can be approximated as,
\begin{eqnarray}\label{eq:Kasner_schw_metric}
   ds^{2} \simeq  \frac{2M}{r}dt^{2} - \frac{r}{2M}dr^{2} + r^{2}d\mathbb{S}_{2} 
\end{eqnarray}

Where $d\mathbb{S}_{2}=d\theta^{2}+ \sin^{2}\theta \  d\phi^{2}$. Note that $(t,r)$ switches roles between timelike coordinate and spacelike coordinate, with $\chi:=t$ (Schwarzschild time) now a spacelike coordinate and $r$ can be parametrized by proper time along the radial infalling timelike geodesics using,

\begin{eqnarray}
    r(\tau_{\circ}) = \left(\frac{3}{2}\sqrt{2M}\tau_{\circ}\right)^{2/3}
\end{eqnarray}

Where $\tau_{\circ}$ is the proper time of a radial infalling observer. Therefore Schwarzschild metric near the singularity can be effectively written in Kasner form with exponents $(2/3,2/3,-1/3)$

\begin{eqnarray}
 ds^{2} \simeq -d\tau^{2}_{\circ} + (\tau_{\circ}/2M)^{-2/3}d\chi^{2} + (2M)^{2/3}\tau_{\circ}^{4/3}d\mathbb{S}_{2} 
\end{eqnarray}

As singularity is approached $\tau_{\circ} \to 0$, the 2-sphere shrinks and locally any 2-sphere is flat in Riemann normal coordinates, thus $d\mathbb{S}_{2} = dy^{2} + dx^{2} + \mathcal{O}(x^{2},y^{2})$.

\begin{eqnarray}
  ds^{2} = - d\tau_{\circ}^{2} + (\tau_{\circ}/2M)^{-2/3}d\chi^{2} + (\tau_{\circ}/\sqrt{2M})^{4/3} (dy^{2} + dx^{2}) 
\end{eqnarray}

therefore, Schwarzschild interior \textit{asymptotically} approaches the Kasner universe in the limit $\tau_\circ \to 0$ and small angular region.
%%%%%%%%%%%%%%%%%%%%%%%%%%%%%%%%%%%%%%%%%%%%%%%%%%%%%%%%%%%%%%%%%%%%%%%%%%%%%%%%%%%%%%%%%%%%%%%%%

%%%%%%%%%%%%%%%%%%%%%%%%%%%%%%%%%%%%%%%%%%%%%%%%%%%%%%%%%%%%%%%%%%%%%%%%
\section*{Acknowledgements}
Mayank thanks the Indian Institute of
Technology (IIT) Madras and the Ministry of Human
Resources and Development (MHRD), India for financial
support.

%\newpage
\appendix

%%%%%%%%%%%%%%%%%%%%%%%%%%%%%%%%%%%%%%%%%%%%%%%%%%%%%%%%%%%%%%%%%%%%%%%%%
%

%%%%%%%%%%%%%%%%%%%%%%%%%%%%%%%%%%%%%%%%

%%%%%%%%%%%%%%%%%%%%%%%%%%%%%%%%%%%%%%%%

\end{document}